\documentclass[10pt,twocolumn,twoside]{IEEEtran}

\usepackage{graphicx}

\usepackage{ifpdf}
\usepackage{cite}

\usepackage[cmex10]{amsmath}
\ifCLASSOPTIONcompsoc
  \usepackage[caption=false,font=normalsize,labelfont=sf,textfont=sf]{subfig}
\else
  \usepackage[caption=false,font=footnotesize]{subfig}
\fi

\usepackage{fixltx2e}
\usepackage{url}
\begin{document}

\title{Nonlinear Blind Source Separation Using Sensor-Independent Signal Representations}

\author{David~N.~Levin%
\thanks{D. Levin is with the Department of Radiology and the Committee on Medical Physics, University of Chicago. Mailing address: 1310 N. Ritchie Ct., Unit 26 AD, Chicago, IL 60610. Telephone: 312-753-9274. Fax: 312-482-8624. Email: d-levin@uchicago.edu}%
\thanks{}}

\markboth{Nonlinear Blind Source Separation}%
{Levin : Blind Source Separation}

\maketitle

\begin{abstract}
Consider a time series of signal measurements $x(t)$, having components $x_k \mbox{ for } k = 1,2, \ldots ,N$. This paper shows how to determine if these signals are equal to linear or nonlinear mixtures of the state variables of two or more statistically-independent subsystems. First, the local distribution of measurement velocities ($\dot{x}$) is processed in order to derive $N$ local vectors at each $x$. If the data are separable, each of these vectors is directed along a subspace traversed by varying the state variable of one subsystem, while all other subsystems are kept constant. Because of this property, these vectors can be used to determine if the data are separable, and, if they are, $x(t)$ can be transformed into a separable coordinate system in order to recover the time series of the independent subsystems.  The method is illustrated by using it to blindly recover the separate utterances of two speakers from nonlinear combinations of their waveforms.
\end{abstract}


%\begin{center} \bfseries EDICS Category: ??? \end{center}

\section{Introduction}

The signals from a process of interest are often contaminated by signals from "noise" processes, which are thought to be statistically independent of the process of interest but are otherwise unknown. This raises the question: can one use the signals to determine if two or more independent processes are present, and, if so, can one derive a representation of the evolution of each of them? In other words, if a system is effectively evolving in a closed box, can one ``explore" the signals emanating from the box in order to learn the number and nature of the subsystems within it? There is a variety of methods for solving this blind source separation (BSS) problem for the special case in which the signals are linearly related to the system states. However, some observed signals (e.g., from biological or economic systems) may be nonlinear functions of the underlying system states, and computational methods for performing nonlinear BSS are limited (\cite{Jutten, Almeida}), even though humans can often perform it quite effortlessly.

Consider an evolving physical system that is being observed by making time-dependent measurements, $x_{k}(t) \mbox{ for } k = 1,2, \ldots ,N$, which are invertibly related to the system's state variables. In Conclusion, we describe how to choose measurements that have this invertibility property. The objective of BSS is to determine if the measurements are mixtures of the state variables of statistically independent subsytems. Specifically, we want to know if there is an invertible, possibly nonlinear, $N \mbox{-component}$ ''mixing" function, $f$, that transforms the measurement time series into a time series of separable states:
\begin{equation}
\label{mixture}
s(t) = f[x(t)] ,
\end{equation}
where $s(t)$ denotes a set of state components, $s_{k}(t) \mbox{ for } k = 1,2, \ldots ,N$, which can be partitioned into two or more statistically independent groups. Note that the mixing function defines a transformation between two coordinate systems on state space: the coordinate system ($x$) defined by the choice of measurements and the coordinate system ($s$) of separable state components.

The method proposed in this paper utilizes a criterion for statistical independence that differs from the conventional one.  Let $\rho_S(s)$ be the probability density function (PDF), defined so that $\rho_S(s) ds$ is the fraction of total time that the trajectory $s(t)$ is located within the volume element $ds$ at location $s$. In the usual formulation of the BSS problem, the mixing function must transform the measurements so that $\rho_S(s)$ is the product of the density functions of individual components (or groups of components)
\begin{equation}
\label{state space factorization}
\rho_S(s) = \prod_{a=1,2, \ldots}{\rho_{a}(s_{(a)})} ,
\end{equation}
where $s_{(a)}$ is a subsystem state variable, comprised of one or more of the components $s_k$. In every formulation of BSS, multiple solutions can be created by permuting the subsystem state variables and/or transforming their components. However, the criterion in (\ref{state space factorization}) is so weak that it suffers from a much worse non-uniqueness problem: namely, solutions can be created by mixing the state variables of other solutions (see~\cite{Hyvarinen-uniqueness} and references therein).

In this paper, the issue of non-uniqueness is circumvented by considering the data's trajectory in $(s,\dot{s}) \mbox{-space}$ ($\dot{s}~=~ds/dt$), instead of $s \mbox{-space}$ (i.e., state space). First, let $\rho_S(s,\dot{s})$ be the PDF in this space, defined so that $\rho_S(s,\dot{s}) ds d\dot{s}$ is the fraction of total time that the location and velocity of $s(t)$ are within the volume element $ds d\dot{s}$ at location $(s,\dot{s})$. As in an earlier paper~\cite{Levin-bss-JAP}, the mixing function must transform the measurements so that $\rho_S(s,\dot{s})$ is the product of the density functions of individual components (or groups of components)
\begin{equation}
\label{phase space factorization}
\rho_S(s,\dot{s}) = \prod_{a=1,2, \ldots}{\rho_{a}(s_{(a)},\dot{s}_{(a)})} .
\end{equation}
Separability in $(s,\dot{s}) \mbox{-space}$ is a stronger requirement than separability in state space. To see this, note that (\ref{state space factorization}) can be recovered by integrating both sides of (\ref{phase space factorization}) over all velocities, but the latter equation cannot be deduced from the former one. In fact, it can be shown that (\ref{phase space factorization}) is strong enough to guarantee that the BSS problem has a unique solution, up to permutations of subsystem state variables and transformations of their components (\cite{Levin-bss-JAP}), and that is why it is studied here.

Because this paper utilizes the separability criterion in (\ref{phase space factorization}), it is fundamentally different from most other BSS techniques, which utilize the weaker criterion in (\ref{state space factorization}). Furthermore, the new method exploits statistical constraints on the measurement velocities in each \textit{local} region of state space. In contrast, existing methods of using velocity information utilize weaker constraints on the \textit{global} distribution of measurement velocities (\cite{Lagrange}).

The new method should be compared to two earlier techniques of performing BSS according to the criterion in (\ref{phase space factorization}). In \cite{Levin-bss-JAP} it was shown that the local second-order correlation matrix of measurement velocity can be taken to define a Riemannian metric on the space of measurements ($x$). Nonlinear BSS can then be performed by finding the transformation to another ($s$) coordinate system, in which this metric is block-diagonal everywhere. In order to construct this new coordinate system, it is necessary to compute the metric's first and second derivatives with respect to $x$. This approach suffers from a practical difficulty: namely, a great deal of data is required to cover the measurement manifold densely enough in order to calculate these derivatives accurately. In contrast, the method proposed in this paper only depends on the computation of the second and fourth-order correlations of the local $\dot{x}$ distribution. This can be done with much less data.

Reference \cite{Levin-IEEE-Trans} describes a second way of performing nonlinear BSS according to the criterion in (\ref{phase space factorization}). Higher-order local correlations of the data's velocity (typically, at least fifth-order correlations) are used to compute multiple scalar quantities (typically, at least six scalars) at each point $x$. A necessary consequence of separability is that certain subgroups of these scalars must map $x \mbox{-space} $ onto a subspace of lower dimensions. Typically, this approach requires more data than the proposed technique because it is necessary to compute local velocity correlations of order greater than four.

The next section describes how to determine if the data are separable and, if so, how to recover a representation of the evolution of each independent subsystem. Section III illustrates the method by using it to recover the utterances of two speakers from unknown nonlinear mixtures of their waveforms.  The last section discusses the implications of this approach.

\begin{figure*}[!tbp]
\begin{center}
\includegraphics[trim=0 2cm 0 1cm,clip,width=6.5in]{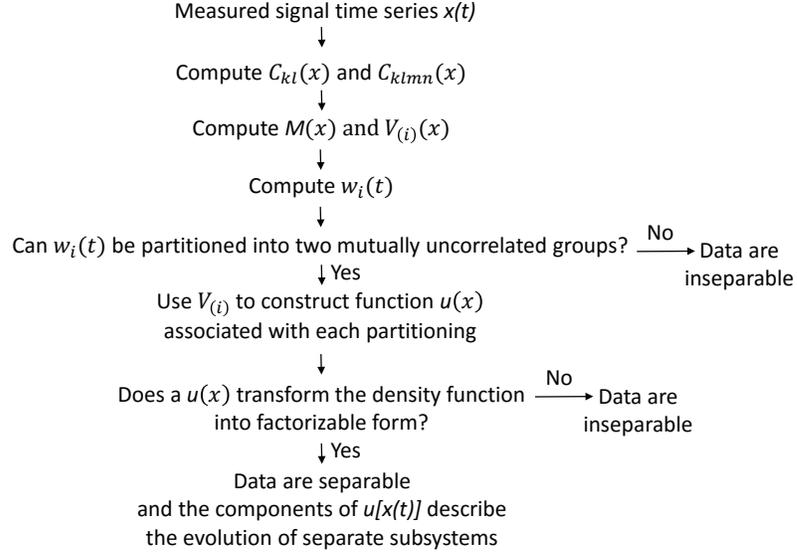}
\caption{ The proposed method of nonlinear blind source separation}
\label{figure1}
\end{center}
\end{figure*}

\section{Method}

This paragraph and Figure \ref{figure1} provide a brief description of the proposed BSS procedure. The procedure is initiated by using the local distribution of measurement velocities to construct $N$ local vectors ($V_{(i)}(x) \mbox{ for } i = 1,2, \ldots ,N$) at each point $x$. If the data are separable, there is a special $s$ coordinate system in which (\ref{phase space factorization}) is true. In that coordinate system it is evident that each vector is directed along the subspace created by varying one subsystem's state variables, while all other subsystems are held constant. Because of this property, these vectors can be used to construct a finite set of functions on $x \mbox{-space}$, and one of those functions must transform the data to a separable coordinate system if it exists. Therefore, we can determine whether the data are separable by seeing if any of these functions transforms the data's density function (or correlations) into a factorizable form. If the data are separable, they can be transformed into the separable coordinate system in order to recover the time course of each subsystem (up to an arbitrary transformations on the state space of each subsystem).
 
The first step is to construct second-order and fourth-order local correlations of the data's velocity
\begin{equation}
\label{C2 definition}
C_{kl}(x) = \, \langle (\dot{x}_k-\bar{\dot{x}}_k) (\dot{x}_l-\bar{\dot{x}}_l) \rangle_{x} 
\end{equation}
\begin{equation}
\label{C4 definition}
\begin{split}
C_{klmn}(x) = \, \langle (\dot{x}_k-\bar{\dot{x}}_k) & (\dot{x}_l-\bar{\dot{x}}_l) \\ 
& (\dot{x}_m-\bar{\dot{x}}_m) (\dot{x}_n-\bar{\dot{x}}_n) \rangle_{x}
\end{split}
\end{equation}
where $\bar{\dot{x}} = \langle \dot{x} \rangle_x$, where the bracket denotes the time average over the trajectory's segments in a small neighborhood of $x$, and where all indices are integers between $1$ and $N$. Because $\dot{x}$ is a contravariant vector, $C_{kl}(x)$ and $C_{klmn}(x)$ are local contravariant tensors of second and fourth rank, respectively. The definition of the PDF implies that $C_{kl}(x)$ and $C_{klmn}(x)$ are two of its moments; e.g.,
\begin{equation}
\label{PDF moment}
C_{kl \ldots}(x) = \frac {\int \rho(x,\dot{x}) (\dot{x}_k-\bar{\dot{x}}_k) (\dot{x}_l-\bar{\dot{x}}_l) \ldots d\dot{x}} {\int \rho(x,\dot{x}) d\dot{x}} ,
\end{equation}
where $\rho(x,\dot{x})$ is the PDF in the $x$ coordinate system, where ``$\ldots$" denotes possible additional indices on the left side and corresponding factors of $\dot{x}-\bar{\dot{x}}$ on the right side, and where all indices are integers between $1$ and $N$. Although (\ref{PDF moment}) is useful in a formal sense, in practical applications all required correlation functions can be computed directly from local time averages of the data (i.e., (\ref{C2 definition})-(\ref{C4 definition})), without explicitly computing the data's PDF. Also, note that velocity ``correlations" with a single subscript vanish identically
\begin{equation}
\label{C_k=0}
C_k(x)=0  .
\end{equation}

Next, let $M(x)$ be any local $N \times N$ matrix, and use it to define $M \mbox{-transformed}$ velocity correlations
\begin{equation}
\label{I2 definition}
I_{kl}(x) = \sum_{1 \leq k', \, l' \leq N} M_{kk'}(x) M_{ll'}(x) C_{k'l'}(x) ,
\end{equation}
\begin{equation}
\label{I4 definition}
\begin{split}
I_{klmn}(x) = \sum_{1 \leq k', \, l', \, m', \, n' \leq N} & M_{kk'}(x) M_{ll'}(x) \\
& M_{mm'}(x) M_{nn'}(x) C_{k'l'm'n'}(x) .
\end{split}
\end{equation}
Because $C_{kl}(x)$ is generically positive definite at any point $x$, it is possible to find a particular form of $M(x)$ that satisfies
\begin{equation}
\label{M definition 1}
I_{kl}(x) = \delta_{kl}
\end{equation}
\begin{equation}
\label{M definition 2}
\sum_{1 \leq m \leq N} I_{klmm}(x) = D_{kl}(x) , 
\end{equation}
where $D(x)$ is a diagonal $N \, x \, N$ matrix. Such an $M(x)$ can be constructed from the product of three matrices: 1) a rotation that diagonalizes $C_{kl}(x)$, 2) a diagonal rescaling matrix that transforms this diagonalized correlation into the identity matrix, 3) another rotation that diagonalizes
\begin{displaymath}
\sum_{1 \leq m \leq N} \tilde{C}_{klmm}(x) ,
\end{displaymath} 
where $\tilde{C}_{klmn}(x)$ is the fourth-order velocity correlation ($C_{klmn}(x)$) after it has been transformed by the first rotation and the rescaling matrix.   As long as $D$ is not degenerate, $M(x)$ is unique, up to arbitrary \textit{local} permutations and/or reflections. In almost all applications of interest, the velocity correlations will be continuous functions of $x$. Therefore, in any neighborhood of state space, there will always be a continuous solution for $M(x)$, and this solution is unique, up to arbitrary \textit{global} permutations and/or reflections. 

In any other coordinate system $x'$, the most general solution for $M'$ is given by
\begin{equation}
\label{M'}
M'_{kl}(x') = \sum_{1 \leq m, \, n \leq N} P_{km} M_{mn}(x) \frac{\partial x_n}{ \partial x'_l} (x') ,
\end{equation}
where $M$ is a matrix that satisfies (\ref{M definition 1}) and (\ref{M definition 2}) in the $x$ coordinate system and where $P$ is a product of permutation, reflection, and identity matrices. This can be proven by substituting this equation into the definition of $I'_{kl}(x')$ and $I'_{klmn}(x')$ and by noting that these quantities satisfy (\ref{M definition 1}) and (\ref{M definition 2}) in the $x'$ coordinate system because (\ref{I2 definition})-(\ref{I4 definition}) satisfy them in the $x$ coordinate system. Note that, by construction, $M$ is not singular, and, therefore, it has a non-singular inverse.

Notice that (\ref{M'}) shows that the rows of $M$ transform as local covariant vectors, up to global permutations and/or reflections. Likewise, the same equation implies that the columns of $M^{-1}$ transform as local contravariant vectors (denoted as $V_{(i)}(x) \mbox{ for } i = 1,2, \ldots ,N$), up to global permutations and/or reflections. Because these vectors are linearly independent, the measurement velocity at each time, $\dot{x}(t)$, can be represented by a weighted superposition of them
\begin{equation}
\label{xDot rep}
\dot{x}(t) = \sum_{1 \leq i \leq N} w_{i}(t) V_{(i)}  ,
\end{equation}
where $w_{i}$ are time-dependent weights. Because $\dot{x}$ and $V_{(i)}$ transform as contravariant vectors, the weights $w_{i}$ must transform as scalars; i.e., they are independent of the coordinate system in which they are computed (except for possible permutations and/or reflections). In this sense, the time series of weights comprises an invariant or coordinate-system-independent representation of the system's velocity.

Now, let's imagine the computation of these weights in the separable ($s$) coordinate system, in which the components of $s$ have been partitioned into disjoint blocks corresponding to (possibly multidimensional) subsystem state variables. In this coordinate system, correlations $C_{Skl \ldots}(s)$, having all indices in block $a$, are the correlations between the components of state variable $s_{(a)}$.  Next, construct the block diagonal matrix $M_S(s)$
\begin{equation}
\label{block-diagonal MS}
M_S(s) = \left( \begin{array}{ccc}  
   M_{S1}(s_{(1)}) & 0 & \ldots \\
   0 & M_{S2}(s_{(2)}) & \ldots \\
   \vdots & \vdots & \ddots 
   \end{array} \right) .
\end{equation}
where submatrix $M_{Sa}$ satisfies (\ref{M definition 1}) and (\ref{M definition 2}) for correlations between components of $s_{(a)}$.
It is not difficult to show that $M_S$ satisfies (\ref{M definition 1}) and (\ref{M definition 2}) in the $s$ coordinate system and that it is unique, up to global permutations and/or reflections. To see this, first note that (\ref{PDF moment}), (\ref{phase space factorization}), and (\ref{C_k=0}) imply that velocity correlations in the $s$ coordinate system vanish if their indices contain a solitary index from any one block. It follows that $C_{Skl}(s)$ is block diagonal and that (\ref{block-diagonal MS}) satisfies the constraint (\ref{M definition 1}), because each block of $M_S$ is defined to transform the corresponding block of $C_{Skl}$ into an identity matrix. In order to prove that (\ref{block-diagonal MS}) satisfies (\ref{M definition 2}), substitute it into the definition of
\begin{equation}
\label{ISklmm}
\sum_{1 \leq m \leq N} I_{Sklmm} .
\end{equation}
Then, note that: 1) when $k$ and $l$ belong to different blocks, each term in this sum vanishes because it factorizes into a product of correlations, one of which has a single index; 2) when $k$ and $l$ belong to the same block and are unequal, each term with $m$ in any other block contains a factor equal to $I_{Skl}$, which vanishes as proved above; 3) when $k$ and $l$ belong to the same block and are unequal, the sum over $m$ in the same block vanishes, because each block of $M_S$ is defined to satisfy (\ref{M definition 2}) for the corresponding subsystem.

After transforming (\ref{xDot rep}) into the $s$ coordinate system, it has the form
\begin{equation}
\label{sDot rep}
\dot{s}(t) = \sum_{1 \leq i, \, j \leq N} w_{i}(t) P_{ij} V_{S(j)}  ,
\end{equation}
where $V_{S(j)}$ is $V_{(j)}$ in the $s$ coordinate system and $P$ is a possible permutation and/or reflection.  Note that $V_{S(i)}$ is the $i^{th}$ column of $M^{-1}_S$. In other words, the $V_{S(i)}$ are the local vectors, which are derived from the local distribution of $\dot{s}$ in the same way that the $V_{(i)}$ were derived from the local distribution of $\dot{x}$. So, (\ref{sDot rep}) shows that the weights that represent $\dot{s}$ are the same (up to a possible permutation and/or reflection) as those that represent $\dot{x}$. 

Observe that each vector $V_{S(i)}$ vanishes except where it passes through one of the blocks of $M^{-1}_S$. Therefore, equation (\ref{sDot rep}) is equivalent to a group of equations, which are formed by projecting it onto each block corresponding to a subsystem state variable. For example, projecting both sides of (\ref{sDot rep}) onto block $a$ gives the result
\begin{equation}
\label{s(a)Dot rep}
\dot{s}_{(a)}(t) = \sum_{\substack{1 \leq i \leq N \\ j \, \in \, block \, a}} w_{i}(t) P_{ij} V_{S(ja)}  .
\end{equation}
Here, $V_{S(ja)}$ is the projection of $V_{S(j)}$ onto block $a$; i.e., it is the column of $M_{Sa}^{-1}$ that coincides with column $j$ of $M^{-1}_S$, as it passes through block $a$. This means that the vectors $V_{S(ja)}$, for $j \in block \, a$, are the local vectors on the $s_{(a)}$ manifold, which are derived from the local distribution of $\dot{s}_{(a)}$ in the same way that the $V_{(i)}$ were derived from the local distribution of $\dot{x}$. Therefore, (\ref{s(a)Dot rep}) shows that each weight $w_{i}(t)$, appearing in the invariant representation of $\dot{x}$, is one of the weights in the invariant representation of the velocity of an isolated subsystem. Notice that each time-dependent weight, $w_{i}(t)$, reflects the evolution of \emph{just one} subsystem; it does not contain information about the evolution of several subsystems. Equation (\ref{s(a)Dot rep}) also shows that the time-dependent weights can be partitioned into subsets such as
\begin{equation} 
\sum_{1 \leq i \leq N} w_{i} P_{ij}
\end{equation}
for $j \in block \, a$, each of which describes the canonical representation of one subsystem's velocity. Each of these subsets of time-dependent weights is statistically independent of the other subsets, and the weights within any of these subsets will usually be correlated with one another.

Because of the block-diagonality of $M$ in the $s$ coordinate system, it is apparent that each vector $V_{S(i)}$ is directed along a subspace traversed by varying the state variables of one subsystem, while all other subsystems are held constant. It follows that the vectors $V_{(i)}$ are directed along the images of those subspaces in the $x$ coordinate system, and this property is heavily exploited below. It is interesting that the $V_{(i)}$ would not have this important property if the definition of $M$ (see (\ref{M definition 1}) and (\ref{M definition 2})) was changed by replacing $\sum_{1 \leq m \leq N} I_{klmm}$ with higher order correlations (e.g., $\sum_{1 \leq m, \, n \leq N} I_{klmmnn}$). 

If the data are separable, the weight components can be partitioned into two groups corresponding to independent (possibly multidimensional) subsystems, and these groups will be mutually uncorrelated. Therefore, the next step is to determine if the weight components can be so partitioned. In principle, these groups should be required to factorize the density function of weights in $w \mbox{-space}$ or $(w,\dot{w}) \mbox{-space}$. In practice, one could simply look for groups of weight components that factorize lower order weight correlations.  If it is found that the weight components cannot be partitioned into two mutually uncorrelated groups, the data are inseparable; i.e., there is only one subsystem (the system itself). On the other hand, if the weight components can be so partitioned, the measurements may or may not be separable. In the following paragraph: 1) it is assumed that there are one or more ways to partition the weights in this manner; 2) the vectors $V_{(i)}$ are used to derive a function corresponding to each partitioning; 3) it is shown that at least one of these must be a transformation to a separable coordinate system, if it exists. In principle, each of these functions can then be used to transform $x(t)$ into the corresponding coordinate system, and the factorizability of the density function of the transformed data can be determined. In practice, it may suffice to determine the factorizability of lower order correlations of the transformed data.  In any event, the measurements are separable if and only if at least one of the computed functions maps the data's density function into a factorizable form. If the data are found to be separable into two subsystems, the transformed data consists of two time series, each one describing the evolution of one subsystem. Then, the above methodology can be applied to the time series of each subsystem in order to determine if it is separable into even smaller independent subsystems.

As described above, in this paragraph, it is assumed that there are one or more ways to partition the weight components into two uncorrelated groups. Then, this information is used to derive a set of functions, which must include a transformation to a separable coordinate system, if it exists. To do this, suppose that the weights have been partitioned into two groups (having $N_1$ and $N_2$ elements), which correspond to two independent subsystems. Because of the block-diagonal structure of $M_S$ (see (\ref{block-diagonal MS})), the vectors associated with either one of these groups are directed along subspaces created by varying the corresponding subsystem's state variable, while holding all other subsystems constant. These subspaces can be used to construct the mapping between the $x$ coordinate system and a separable coordinate system on state space. First, choose some point $x_0$, and use the vectors in group $1$ to construct an $N_1 \mbox{-dimensional}$ subspace (called the ''first subspace of type $1$"), which contains points that can be reached by starting at $x_0$ and then varying $s_{(1)}$, the state variable of subsystem $1$. Impose on this subspace some smoothly varying coordinates  $u_{(1)}$ (having components $u_{(1)a} \mbox{ for } a = 1,2, \ldots ,N_1$). Next, at each point in this subspace, use the vectors in group $2$ to construct an $N_2 \mbox{-dimensional}$ subspace, which contains points that can be reached by starting at this point and then varying $s_{(2)}$, the state variable of subsystem $2$. Each point in each one of these subspaces is assigned a constant value of $u_{(1)}$: namely, the value of $u_{(1)}$ at the "starting point" in the first subspace of type $1$. In the same way, construct an $N_2 \mbox{-dimensional}$ subspace (called the "first subspace of type $2$), which contains points that can be reached by starting at $x_0$ and then varying $s_{(2)}$. Then, impose on this subspace some smoothly varying coordinate system $u_{(2)}$ (having components $u_{(2)b} \mbox{ for } b = 1,2, \ldots ,N_2$). Finally, at each point in this subspace, use the vectors in group $1$ to construct an $N_1 \mbox{-dimensional}$ subspace, which contains points that can be reached by starting at this point and then varying $s_{(1)}$. Each point in each one of these subspaces is assigned a constant value of $u_{(2)}$: namely, the value of $u_{(2)}$ at the "starting point" in the first subspace of type $2$. In this way, each point in the space is assigned values of both $u_{(1)}$ and $u_{(2)}$, thereby defining a $u = (u_{(1)}, u_{(2)})$ coordinate system. Note that the $u$ coordinate system has been constructed so that subspaces having constant $u_{(1)}$ coincide with subspaces having constant $s_{(1)}$. Likewise, subspaces having constant $u_{(2)}$ coincide with subspaces having constant $s_{(2)}$. This means that the $u$ coordinates are the same as the $s$ coordinates, except for transformations among the components of $s_{(1)}$ and transformations among the components of $s_{(2)}$. Because such subspace-wise transformations do not affect separability, $u$ must be a separable coordinate system, and, by construction, we know the transformation $u(x)$ that maps the measurements onto these separable coordinates. Now, consider the group of functions, which can be derived in this manner from each way of partitioning the weight components. The above result implies that the data are separable if and only if this group contains a transformation to a separable coordinate system.

\begin{figure*}[!tbp]
\centering
\subfloat[]{\includegraphics[width=1.7in]{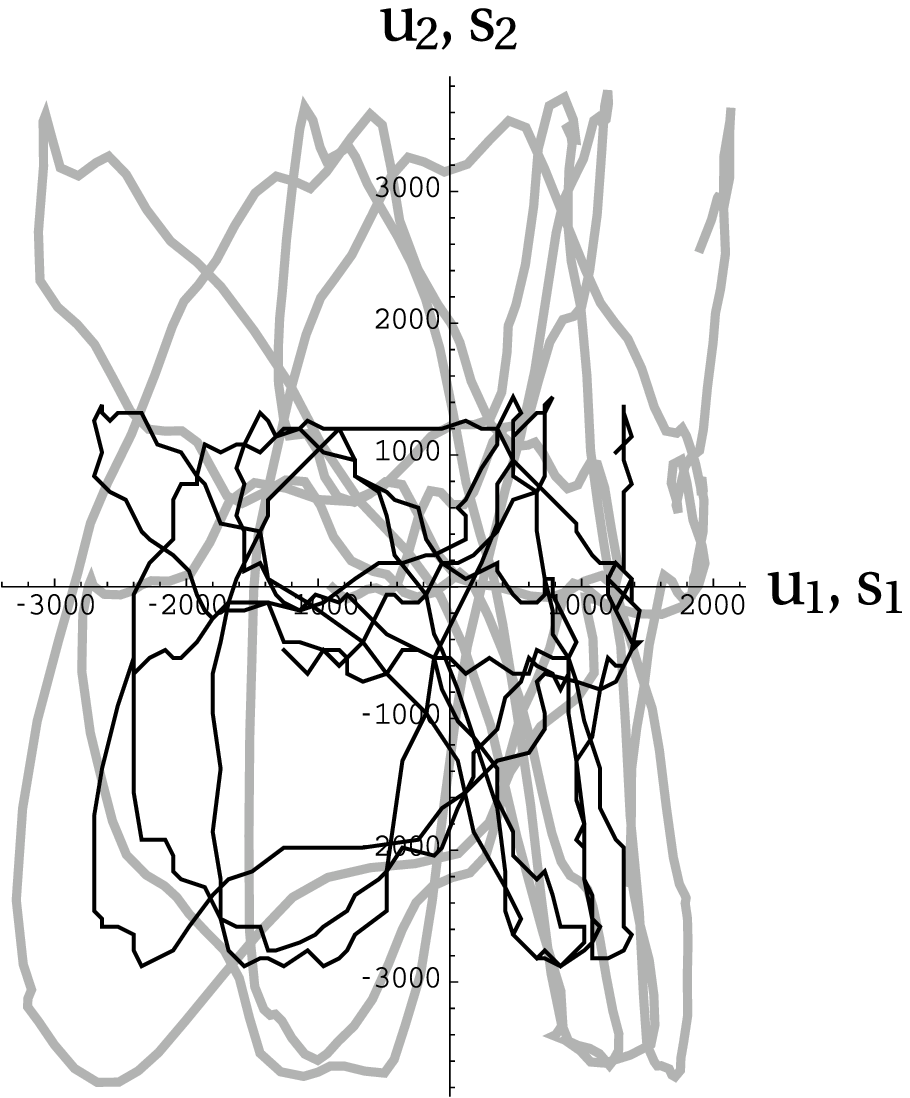}%
\label{fig_s(t)}}
\hfil
\subfloat[]{\includegraphics[width=1.7in]{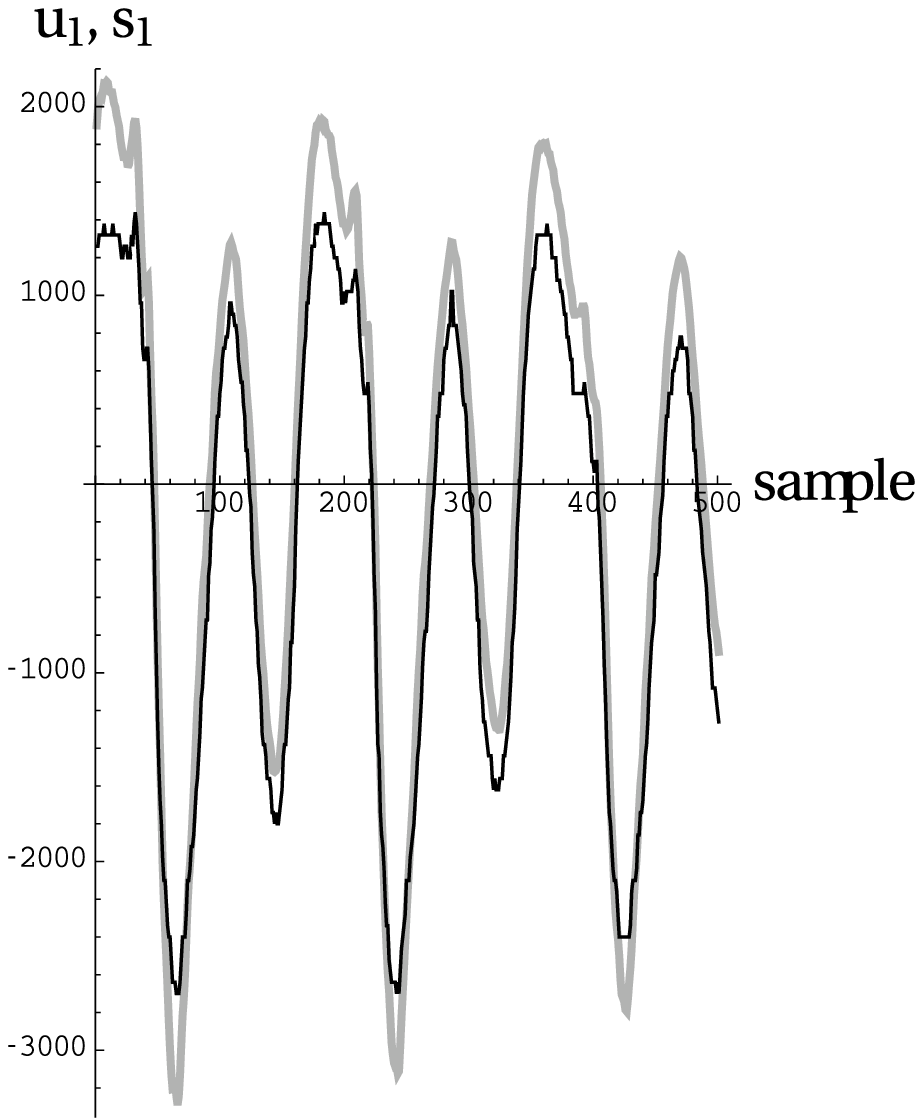}%
\label{fig_s1(t)}}
\hfil
\subfloat[]{\includegraphics[width=1.7in]{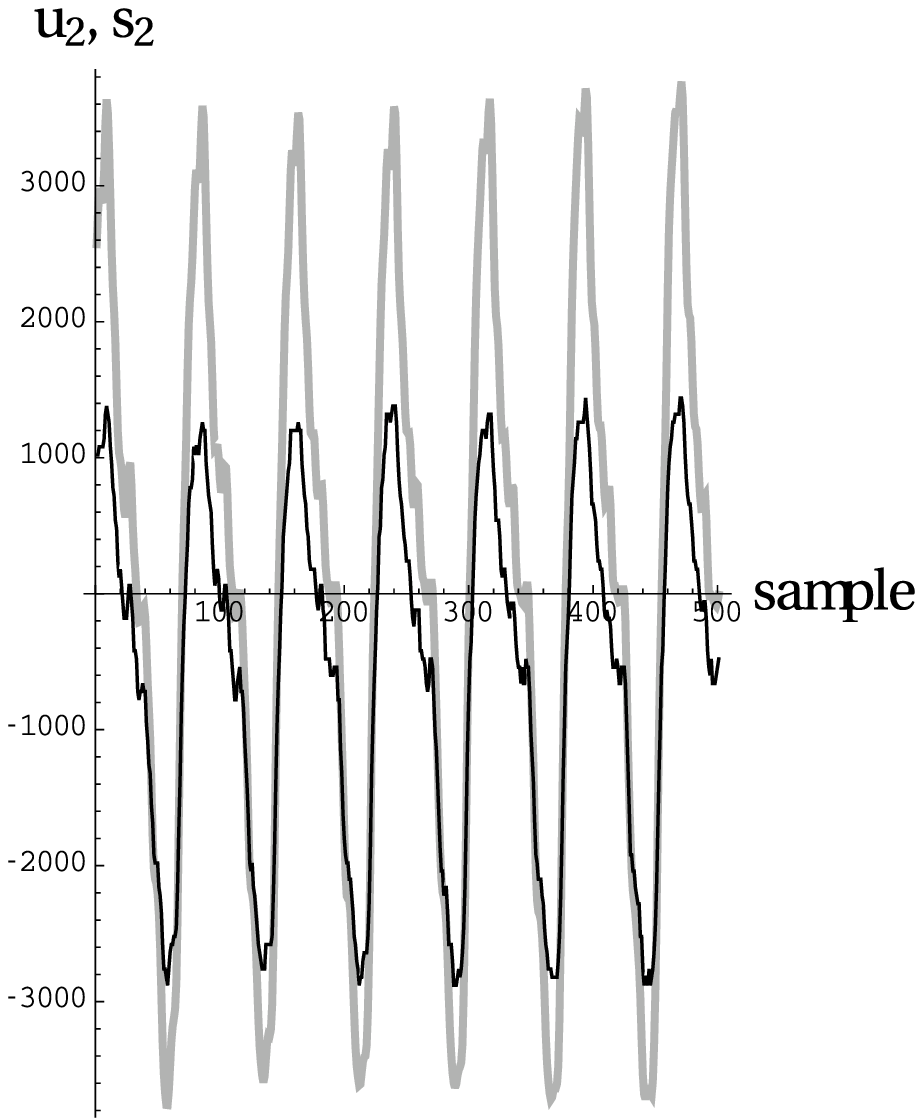}%
\label{fig_s2(t)}}
\caption{ (a) The thick gray line depicts the trajectory of 30 ms of the two speakers' unmixed speech in the $s$ coordinate system, in which each component is equal to one speaker's speech amplitude. The thin black line depicts the waveforms ($u$) of the two speakers during the same time interval, recovered by blindly processing their nonlinearly mixed speech.  Panels (b) and (c) show the time courses of $s_1$ and $u_1$ and of $s_2$ and $u_2$, respectively.}
\label{figure2}
\end{figure*}

\begin{figure*}[!tbp]
\centering
\subfloat[]{\includegraphics[width=1.7in]{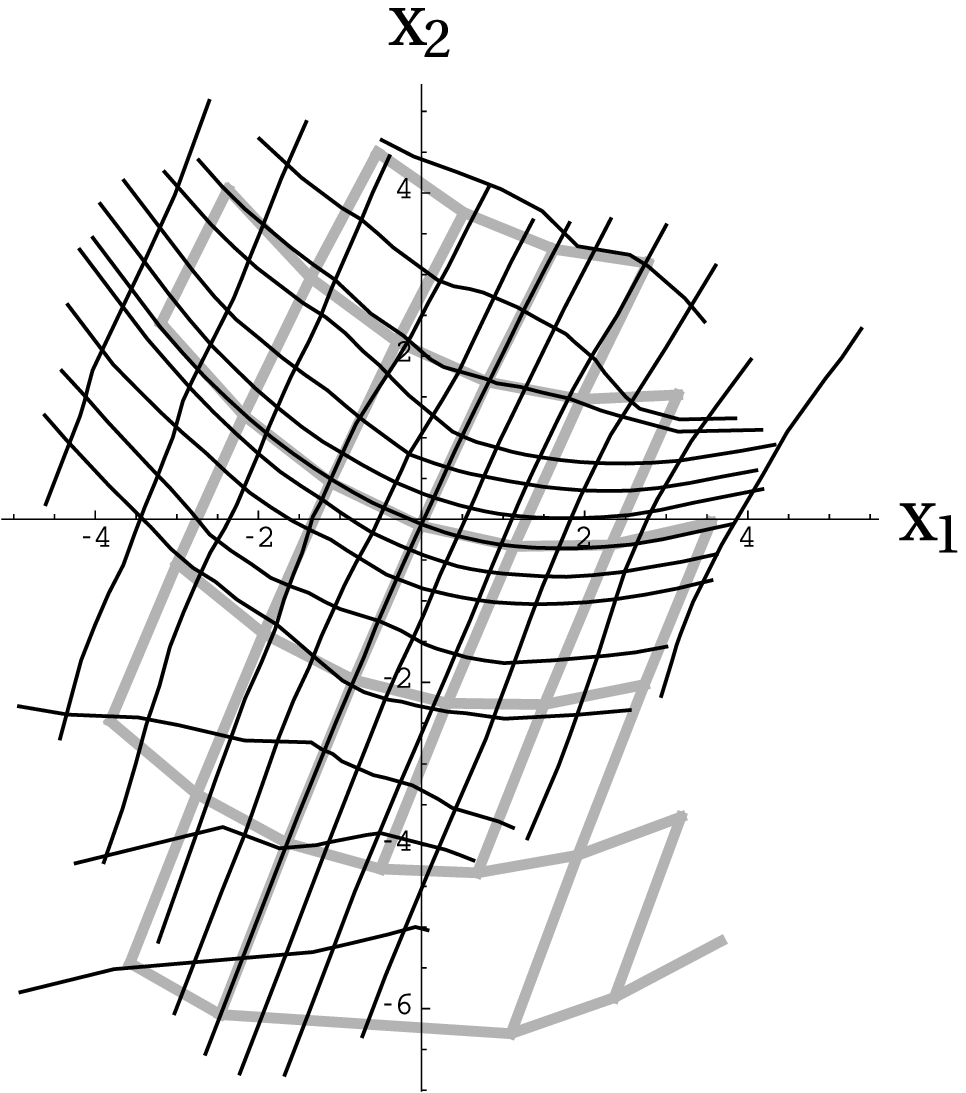}%
\label{fig_warped_grid}}
\hfil
\subfloat[]{\includegraphics[width=1.7in]{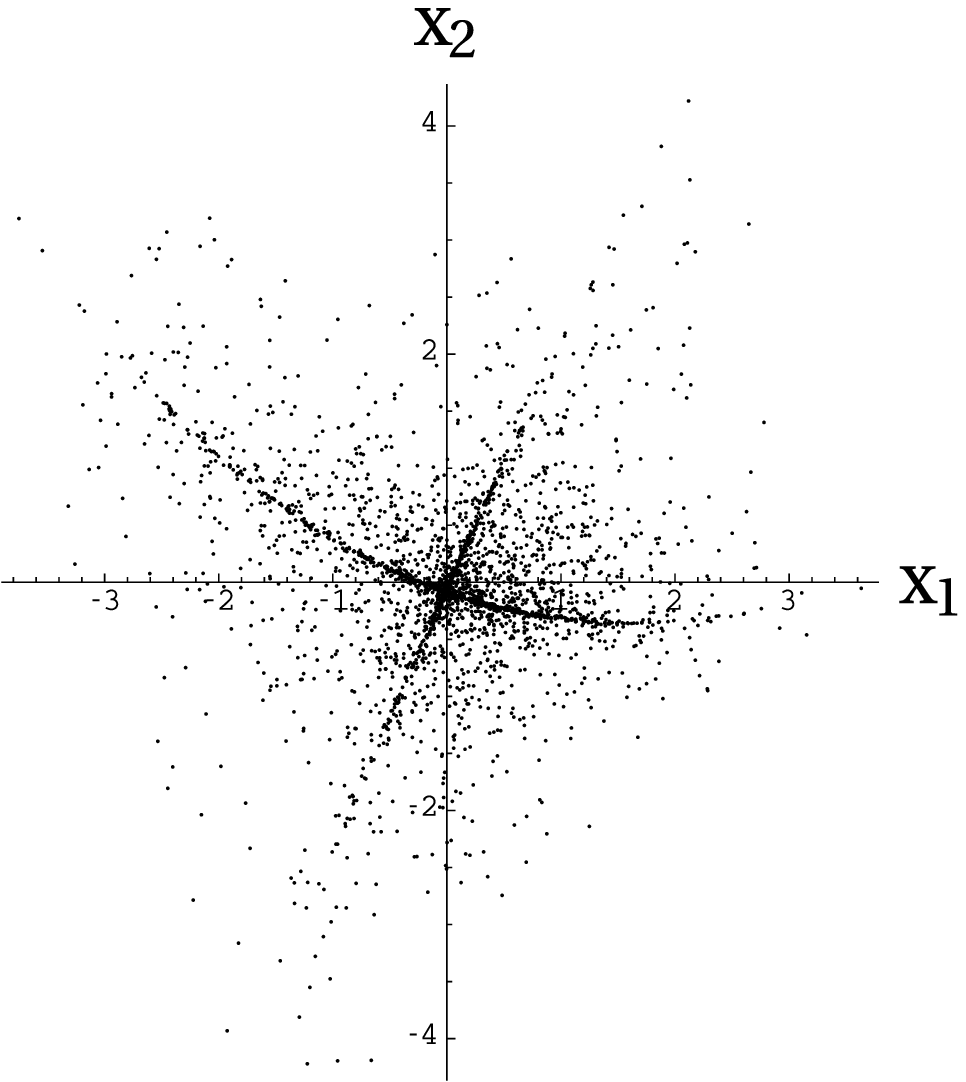}%
\label{fig_measurements}}
\hfil
\subfloat[]{\includegraphics[width=1.7in]{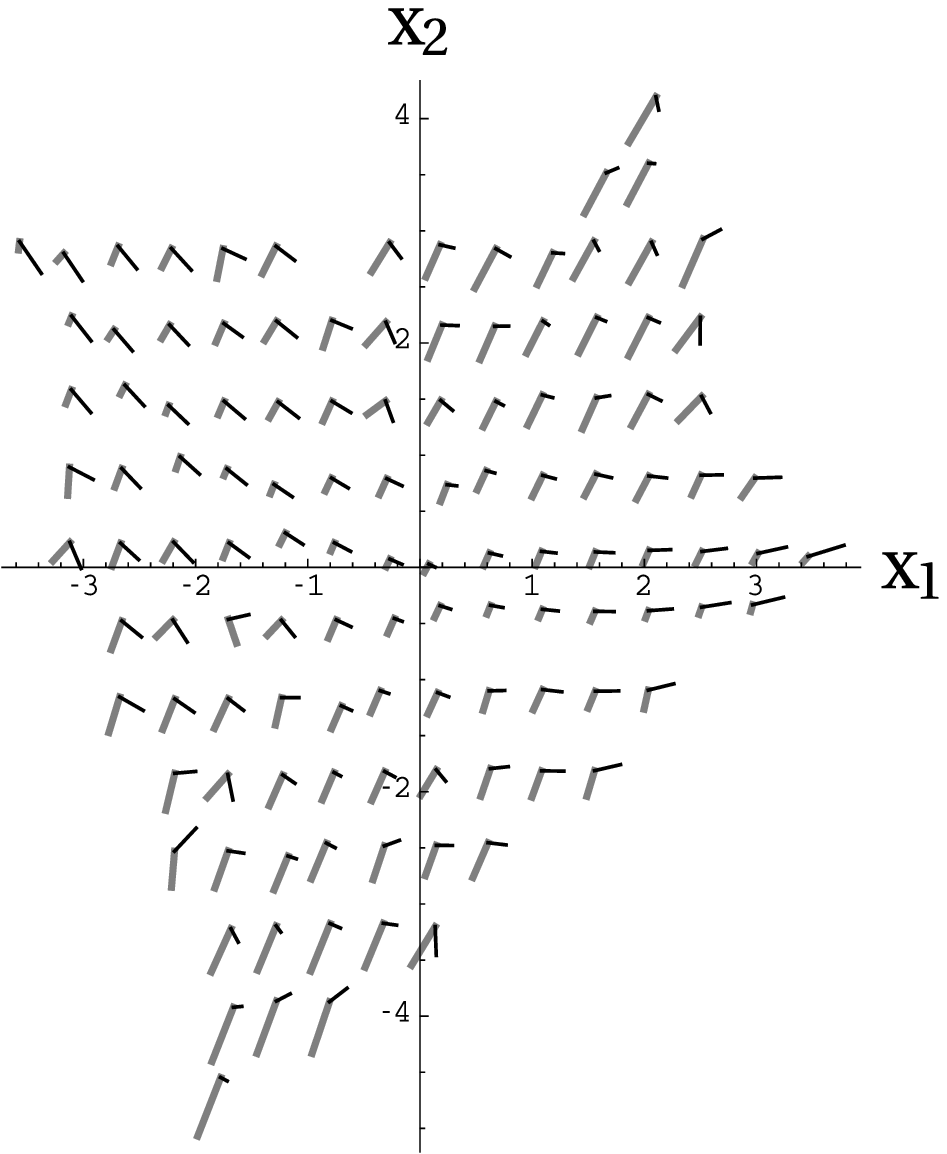}%
\label{fig_V}}
\caption{ (a) The thick gray lines are a regular Cartesian grid of lines with constant $s_1$ and constant $s_2$, after they were nonlinearly mapped into the $x$ coordinate system by the mixing function in (\ref{mixing}). The thin black lines depict lines of constant $u_1$ and $u_2$, where $u$ is a separable coordinate system derived from the measurements. (b) A random subset of the measurements along the trajectory of the mixed waveforms, $x(t)$. (c) The thick gray and thin black lines show the local vectors, $V_{(1)}$ and $V_{(2)}$, respectively, after they have been uniformly scaled for the purpose of display.}
\label{figure3}
\end{figure*}

\begin{figure*}[!tbp]
\centering
\subfloat[]{\includegraphics[width=1.7in]{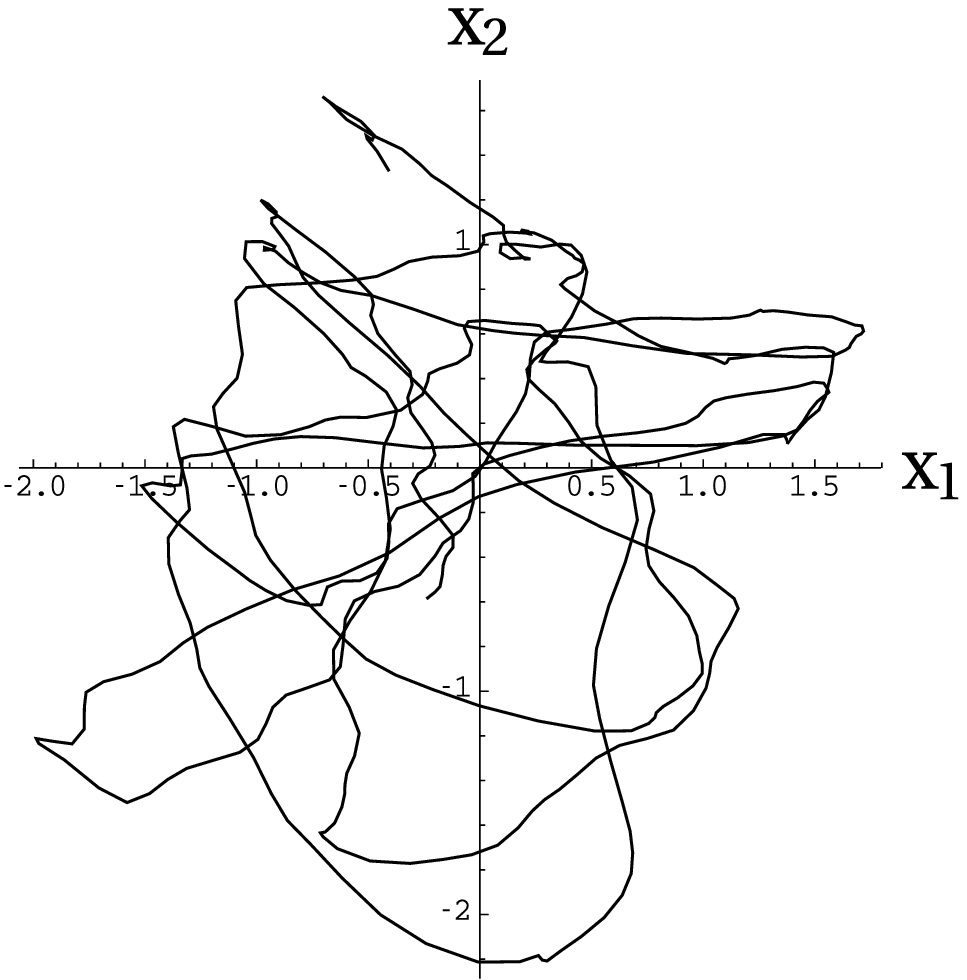}%
\label{fig_x(t)}}
\hfil
\subfloat[]{\includegraphics[width=1.7in]{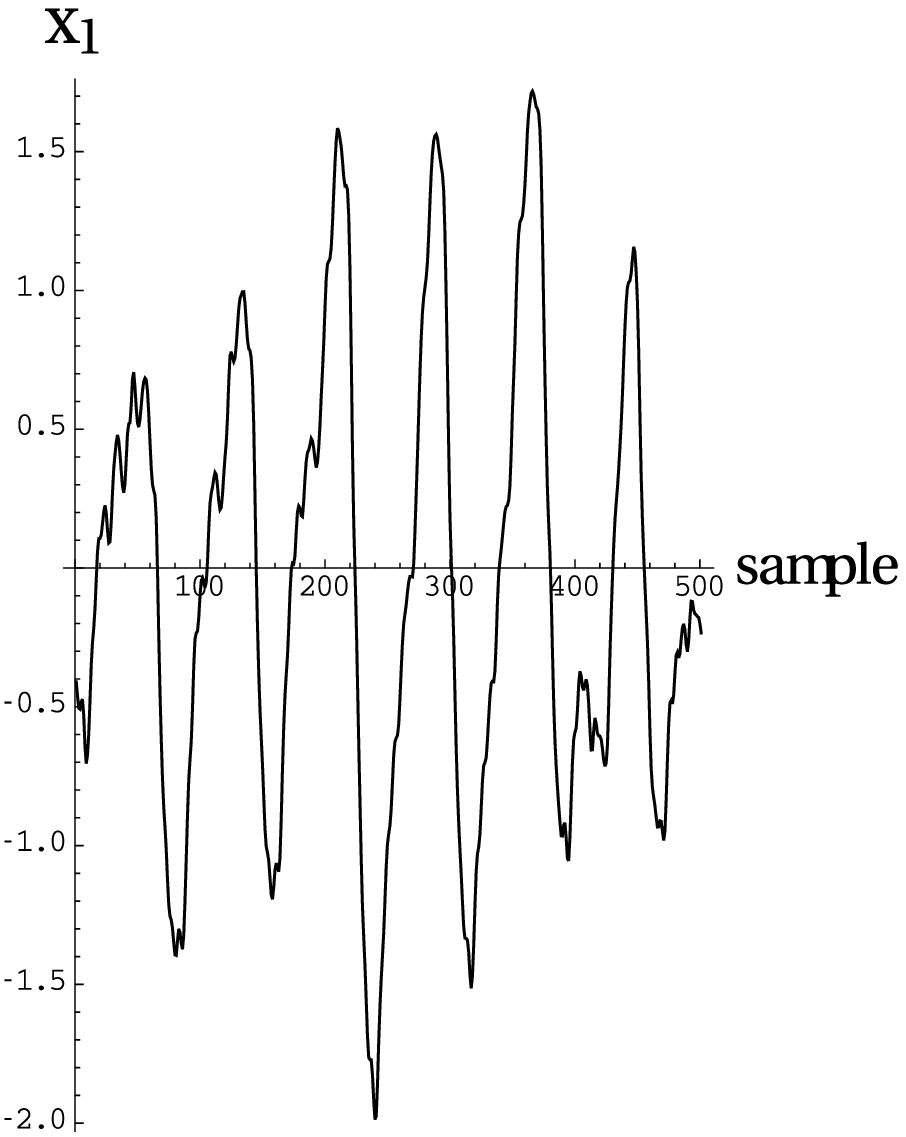}%
\label{fig_x1(t)}}
\hfil
\subfloat[]{\includegraphics[width=1.7in]{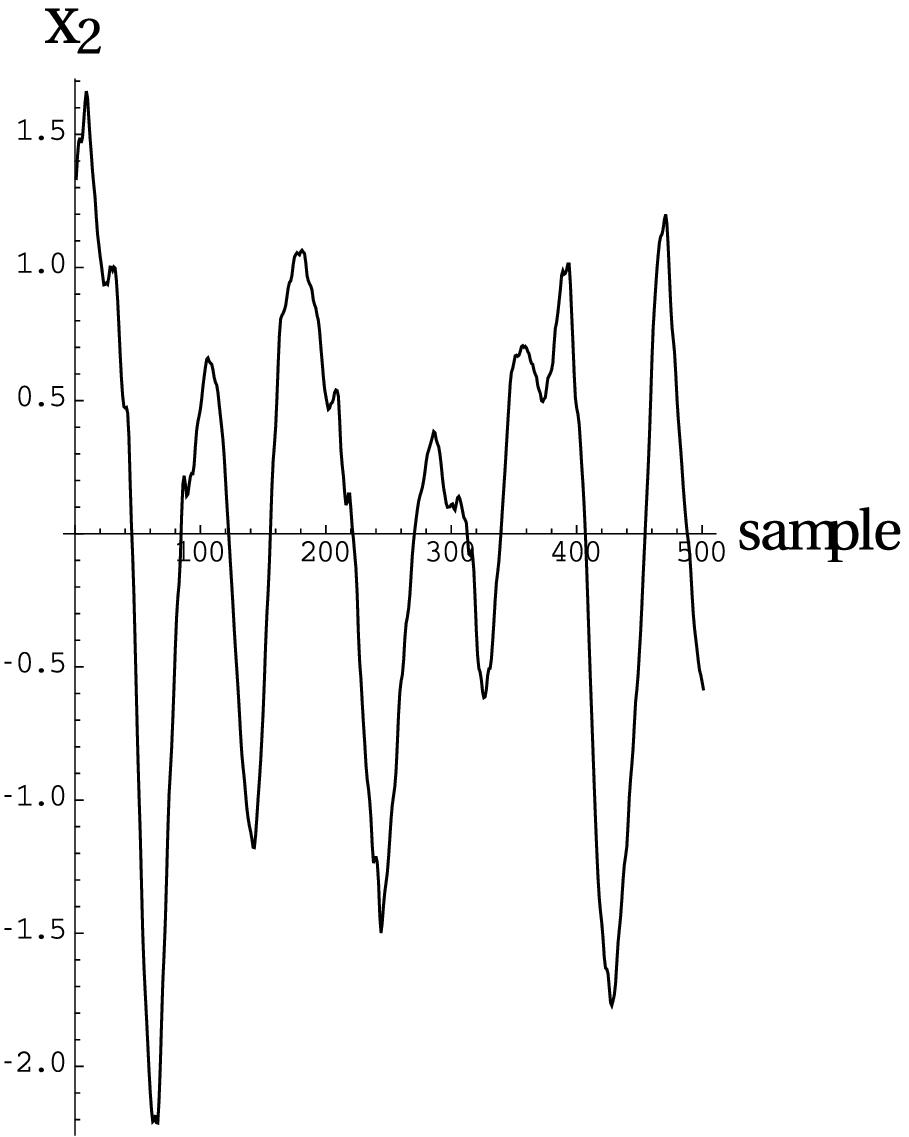}%
\label{fig_x2(t)}}
\caption{(a) The trajectory of measurements, $x(t)$, during the 30 ms time interval depicted in Figure \ref{figure2}.  Panels (b) and (c) show the time courses of $x_1$ and $x_2$, respectively.}
\label{figure4}
\end{figure*}

\begin{figure*}[!tbp]
\centering
\subfloat[]{\includegraphics[width=1.7in]{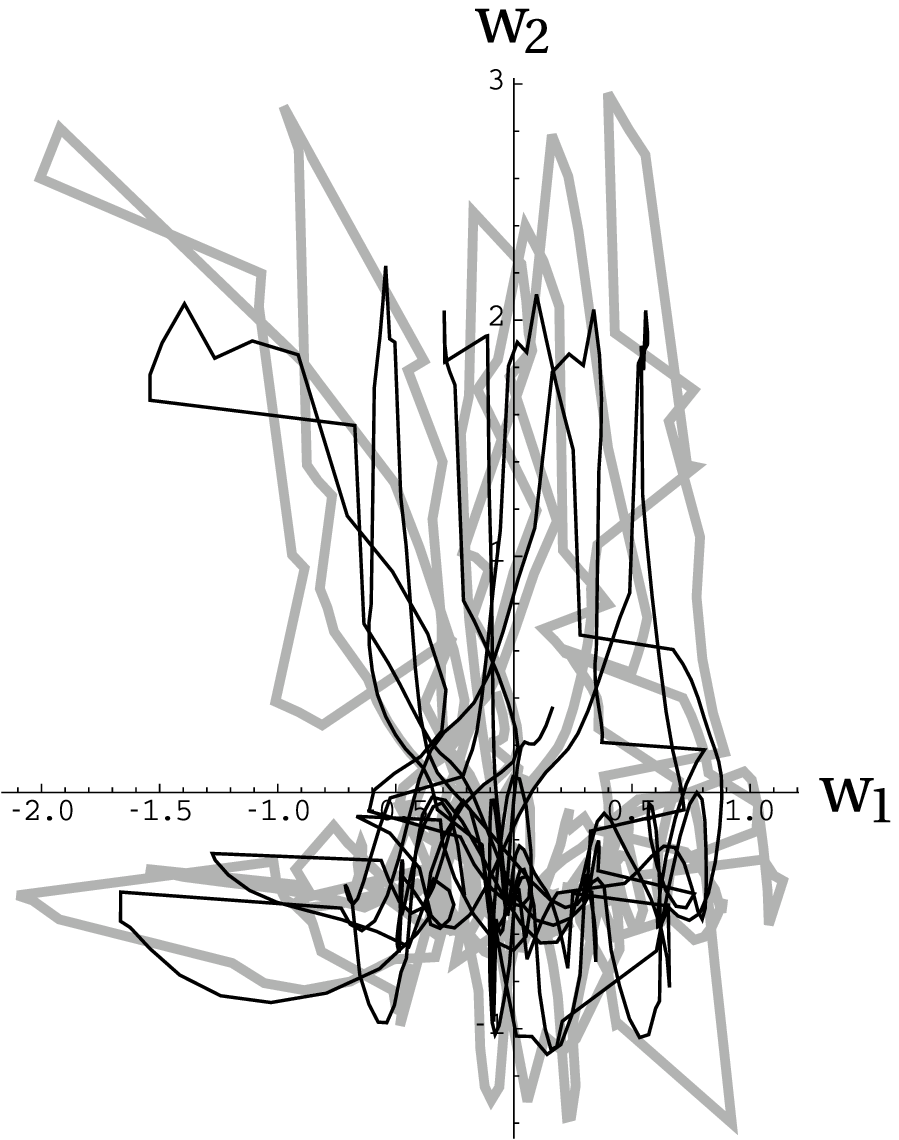}%
\label{fig_W(t)}}
\hfil
\subfloat[]{\includegraphics[width=1.7in]{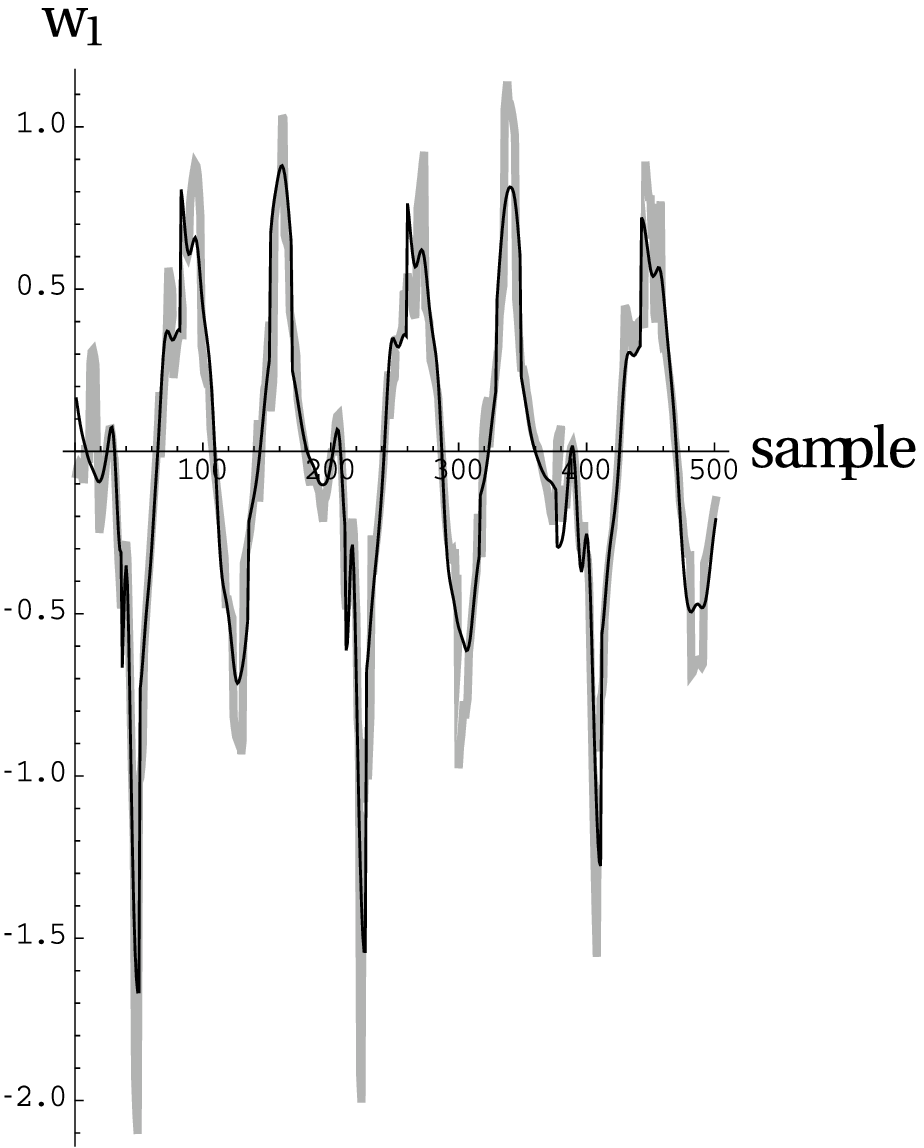}%
\label{fig_W1(t)}}
\hfil
\subfloat[]{\includegraphics[width=1.7in]{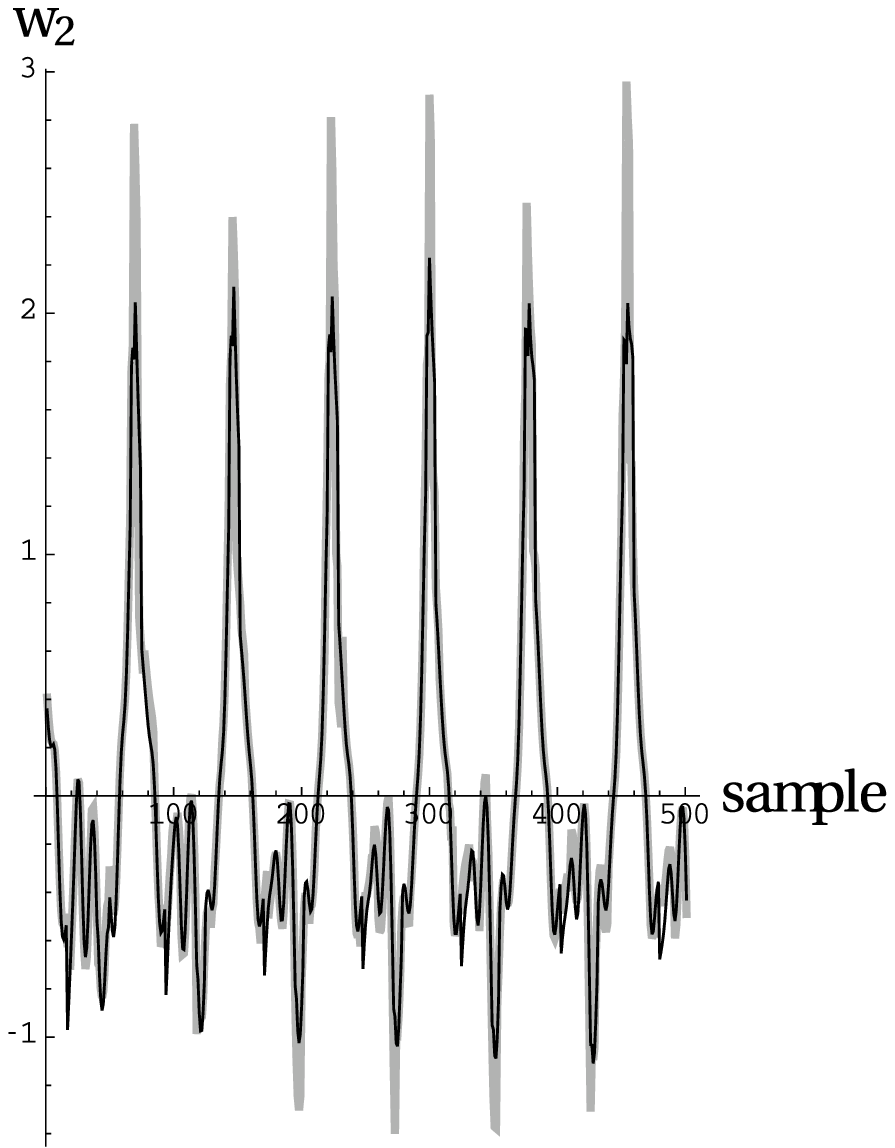}%
\label{fig_W2(t)}}
\caption{The thin black line in panel (a) is the trajectory of the weights, $w_{i}(t)$, which were derived from the measurements in Figure \ref{figure4} and which form a coordinate-system-independent representation of the velocity ($\dot{x}(t)$) of those measurements. The thick gray line in (a) is the time course of the weights, which were derived directly from the unmixed waveforms in Figure \ref{figure2} and which form the coordinate-system-independent representation of $\dot{s}(t)$ during the same time interval.   Panels (b) and (c) show the time courses of the first and second components, respectively, of the lines in panel (a).}
\label{figure5}
\end{figure*}

\section{Experiments}

In this section, the new BSS technique is illustrated by using it to disentangle nonlinear mixtures of the waveforms of two male speakers. Each speaker read an English text, consisting of a thirty-second excerpt from one of two audio book recordings. The waveform of each speaker, $s_{k}(t)$ (for $k=1,2$), was sampled 16,000 times per second with two bytes of depth. The thick gray lines in Figure \ref{figure2} show the two speakers' waveforms during a short (30 ms) interval. These waveforms were then  mixed by the nonlinear functions
\begin{equation}
\label{mixing}
\begin{split}
f_{1}(s) &= 0.763 s_1 + (958 - 0.0225 s_2)^{1.5} \\
f_{2}(s) &= 0.153 s_2 + (3.75 * 10^7-763 s_1 - 229 s_2)^{0.5} ,
\end{split}
\end{equation}
where $-2^{15} \leq s_1, s_2 \leq 2^{15}$. This is one of a variety of nonlinear mixing functions that were tried with similar results. The measurements, $x_{k}(t)$, were taken to be the variance-normalized, principal components of the sampled waveform mixtures, $f_{k}[s(t)]$. Figure \ref{fig_warped_grid} shows how this nonlinear mixing maps an evenly-spaced Cartesian grid in the $s$ coordinate system onto a warped grid in the $x$ coordinate system. Figure \ref{fig_measurements} shows a random subset of the measurements $x(t)$, and Figure \ref{figure4} shows the time course of the measurements $x(t)$ during the same short time interval depicted in Figure \ref{figure2}. When either measured waveform, $x_{1}(t)$ or $x_{2}(t)$, was played as an audio file, it sounded like a confusing superposition of two voices, which were quite difficult to understand. The entire set of 500,000 measurements, consisting of $x$ and $\dot{x}$ at each sampled time, was sorted into a $16 \, \times \, 16$ array of bins. Then, the $\dot{x}$ distribution in each bin was used to compute local velocity correlations (see (\ref{C2 definition}) and (\ref{C4 definition})), and these were used to derive $M$ and $V_{(i)}$ for each bin. Figure \ref{fig_V} shows these local vectors at each point. Finally, (\ref{xDot rep}) was used to transform the measurement velocity ($\dot{x}(t)$) at each time into its coordinate-system-independent representation, given by the weights, $w_{i}(t)$. The thin black lines in Figure \ref{figure5} show the time-dependent weights, which were derived from the waveform mixtures in Figure \ref{figure4}, using (\ref{xDot rep}). The thick gray lines in Figure \ref{figure5} show the time-dependent weights, which were derived directly from each of the unmixed waveforms in Figure \ref{figure2}, using (\ref{s(a)Dot rep}). Notice that the weights, derived from the mixed and unmixed waveforms, are nearly the same, despite the fact that these waveforms differed significantly because of nonlinear mixing (e.g., compare Figures \ref{figure4} and \ref{figure2}). This illustrates the coordinate-system-independence of the weight time series.
 
The correlation between the two weight time series, $w_{1}(t)$ and $w_{2}(t)$, was quite small (namely, -0.0016). In order to determine if these weights correspond to completely separable subsystems, the vectors associated with these two weights were used to construct the putative coordinate transformation, $u(x)$. As described in Method, this was done by using these vectors to construct a family of lines having constant values of $u_1$, together with a family of lines having constant values of $u_2$. The thin black lines in Figure \ref{fig_warped_grid} depict these lines for evenly space values of $u_1$ and $u_2$. The function, $u(x)$, defined by these curves was the only possible transformation to a separable coordinate system. As in Method, the separability of the data could be determined by using this function to transform $x(t)$ into the $u$ coordinate system and by verifying that $u[x(t)]$ has a factorizable density function (or factorizable correlation functions). 

In this illustrative example, the separability of the data in the $u$ coordinate system was verified by showing that the $u$ and $s$ coordinate systems differed by component-wise transformations, which do not affect separability. This was done by comparing the lines of constant $u_1$ and $u_2$ to lines of constant $s_1$ and $s_2$, which are the thin black and thick gray lines, respectively, in Figure \ref{fig_warped_grid}. Notice the near coincidence of the families of lines of constant $u_1$ (or $u_2$) and the families of lines with constant $s_1$ (or $s_2$). This demonstrates that the $u$ and $s$ coordinate systems differ by component-wise transformations of the form: $u = (g(s_{1}), h(s_{2}))$ where $g$ and $h$ are monotonic.  Because the data are separable in the $s$ coordinate system and because component-wise transformations do not affect separability, the data must also be separable in the $u$ coordinate system. Therefore, we have accomplished the objectives of BSS: namely, by blindly processing the data $x(t)$, we have determined that the system is separable, and we have computed the transformation, $u(x)$, to a separable coordinate system.

The transformation $u(x)$ can be applied to the data $x(t)$ to recover the original unmixed waveforms, up to component-wise transformations.  The resulting waveforms, $u_{1}[x(t)]$ and $u_{2}[x(t)]$, are depicted by the thin black lines in Figure \ref{figure2}, which also shows the trajectory of the unmixed waveforms in the $s$ coordinate system. Notice that the two trajectories, $u[x(t)]$ and $s(t)$, are similar except for component-wise transformations along the two axes. The component-wise transformation is especially noticeable as a stretching of $s(t)$ with respect to $u[x(t)]$ along the positive $s_2$ axis. When each of the recovered waveforms, $u_{1}[x(t)]$ and $u_{2}[x(t)]$, was played as an audio file, it sounded like a completely intelligible recording of one of the speakers. In each case, the other speaker was not heard, except for a faint ``buzzing" sound in the background. Therefore, the component-wise transformations, which related the recovered waveforms to the original unmixed waveforms, did not noticeably reduce intelligibility. 

\section{Conclusion}

This paper describes how to determine if time-dependent signal measurements, $x(t)$, are comprised of linear or nonlinear mixtures of the state variables of statistically-independent subsystems. First, the local distribution of measurement velocities, $\dot{x}$, is used to construct local vectors at each point $x$. If the data are separable, each of these vectors is directed along a subspace traversed by varying the state variable of one subsystem, while all other subsystems are kept constant. Because of this property, these vectors can be used to derive a finite set of functions, $u(x)$, which must include the transformation to a separable coordinate system, if it exists. Therefore, separability can be determined by testing the separability of the data, after it has been transformed by each of these mappings. Furthermore, if the data are separable, we can recover a time series that describes the evolution of each subsystem. Therefore, nonlinear blind source separation has been accomplished.

Some comments on this result:
\begin{enumerate}
\item  The time-dependent weights are independent of the choice of sensors used to observe the underlying physical process, and in that sense they comprise an intrinsic or ``inner" property of that process. To see this, note that different sets of sensors detect different mixtures of signals from that process, and different signal mixtures simply describe the system's state in different coordinate systems. Because the weights are scalars (up to permutations and/or reflections), they are independent of the nature of the coordinate system in which they are derived. Therefore, they are also independent of the observer's choice of sensors, as asserted above. This type of sensor-independent signal representation should be contrasted with conventional signal representations, which contain mixtures of information about the sensors, together with information intrinsic to the underlying processes.
\item This paper shows how to perform nonlinear BSS for the case in which the measurements are invertibly related to the state variables of the underlying system. Invertibility can almost be guaranteed by observing the system with a sufficiently large number of independent sensors: specifically, by utilizing at least $2N+1$ independent sensors, where $N$ is the dimension of the system's state space. In this case, the sensors' output lies in an $N \mbox{-dimensional}$ subspace within a space of at least $2N+1$ dimensions. Dimensional reduction techniques (e.g., \cite{Roweis}) can be used to find the subspace coordinates corresponding to the sensor outputs.  Because an embedding theorem asserts that this subspace is very unlikely to self-intersect (\cite{Sauer}), the coordinates on this subspace constitute synthetic ''measurements" that are almost certainly invertibly related to the system's state space. In other words, suitable synthetic measurements can be found by adding more and more sensors until the number of sensors is more than twice the dimensionality ($N$) of the subspace containing the sensors' outputs.
\item Theoretically, the proposed method can be applied to measurements described by any diffeomorphic mixing function. However, in practice, more data will have to be analyzed in order to handle mixing functions with more pronounced nonlinearities. This is because rapidly varying mixing functions may cause the local vectors ($V_{(i)}$) to vary rapidly in the measurement coordinate system, and, therefore, it will be necessary to compute those vectors in numerous small neighborhoods.
\item More data will also be required to apply this method to state spaces having higher dimensions. In Experiments, thirty seconds of data (500,000 samples) were used to recover two waveforms from measurements of two nonlinear mixtures. In other experiments, approximately six minutes of data (6,000,000 samples) were used to cleanly recover the waveforms of four sound sources (two speakers and two piano performances) from four signal mixtures. As expected, blind separation of the 4D state space did require more data, but it was not a prohibitive amount.
\item The method does not require a lot of computational resources. In any event, the most computationally expensive tasks are the binning of the measurement data and the computation of the local vectors, $V_{(i)}$, in each bin. If necessary, these calculations can be parallelized across multiple CPUs.
\end{enumerate}

\section{Acknowledgement}
 The author is grateful to Michael A. Levin for his careful reading of the manuscript and for his suggestions on how to organize the text.

%\section{References}
\bibliographystyle{IEEEtran}

\begin{thebibliography}{1}

\bibitem{Jutten} C.~Jutten and J.~Karhunen, ``Advances in blind source separation (BSS) and independent component analysis (ICA) for nonlinear mixtures," \textit{International J. Neural Systems}, vol. 14, pp. 267-292, 2004.

\bibitem{Almeida} L. Almeida, \textit{Nonlinear Source Separation. Synthesis Lectures on Signal Processing}. Princeton, New Jersey: Morgan and Claypool Publishers, 2005.

\bibitem{Hyvarinen-uniqueness} A.~Hyv\"{a}rinen and P.~Pajunen, ``Nonlinear independent component analysis: existence and uniqueness results," \textit{Neural Networks}, vol. 12, pp. 429-439, 1999.

\bibitem{Levin-bss-JAP}  D.~N. Levin, ``Using state space differential geometry for nonlinear blind source separation," \textit{J. Applied Physics}, vol. 103, art. no. 044906, 2008.

\bibitem{Lagrange} S. Lagrange, L. Jaulin, V. Vigneron, C. Jutten, ``Analytic solution of the blind source separation problem using derivatives," in \textit{Independent Component Analysis and Blind Signal Separation}, LNCS, vol. 3195, C.~G. Puntonet and A.~G. Prieto (eds). Heidelberg: Springer, 2004, pp. 81-88.

\bibitem{Levin-IEEE-Trans}  D.~N. Levin, ``Performing nonlinear blind source separation with signal invariants," \textit {IEEE Trans. Signal Processing}, vol. 58, pp. 2131-2140, 2010.

\bibitem{Roweis} S.~T. Roweis and L.~K.Saul, ``Nonlinear dimensionality reduction by locally linear embedding," \textit{Science}, vol. 290, pp. 2323-2326, 2000.

\bibitem{Sauer} T. Sauer, J.~A. Yorke, M. Casdagli, ``Embedology," \textit{J. Statistical Physics}, vol. 65, pp. 579-616, 1991.


\end{thebibliography}

\begin{IEEEbiography}[{\includegraphics[width=1in,height=1.25in,clip,keepaspectratio]{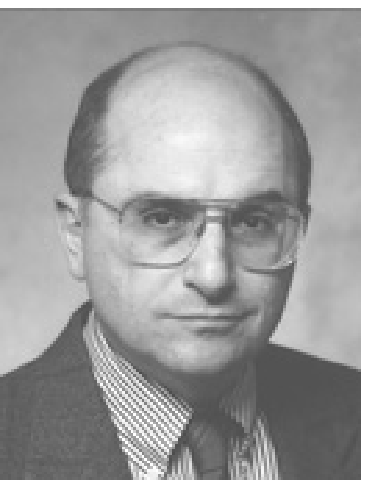}}]{David N. Levin}
 received his Ph.D. in theoretical physics from Harvard University in 1970 and did research in quantum field theory until 1977, when he entered medical school at the University of Chicago.  He joined the faculty after receiving an M.D. and completing radiology residency at the University.  During 1987-1999, he served as Director of Clinical MRI at the University and during 1999-2005 he co-directed the University's Brain Research Imaging Center.  He is currently an emeritus professor in the Department of Radiology.  His past research interests have included the search for all renormalizable quantum field theories, multimodality 3D brain imaging, computer-assisted neurosurgery, image segmentation, methodology for mapping the brain with functional MRI, and the use of prior knowledge of sparse support to increase the speed of MR image acquisition.  His current research is focused on various aspects of signal processing, including nonlinear blind source separation, speech separation, and sensor-independent signal representations.
\end{IEEEbiography}

%\newpage


%\vfill


\end{document}